\documentclass{article}\usepackage{frascatiphys,here,graphicx,subfigure}\begin{document}
\title{STABILITY OF THE SOLUTIONS OF INSTANTANEOUS BETHE--SALPETER EQUATIONS WITH
CONFINING INTERACTIONS}
\author{Wolfgang LUCHA\\{\em Institute for
High Energy Physics, Austrian Academy of Sciences,}\\{\em
Nikolsdorfergasse 18, A-1050 Vienna, Austria}\\Franz
F.~SCH\"OBERL\\ {\em Faculty of Physics, University of
Vienna,}\\{\em Boltzmanngasse 5, A-1090 Vienna, Austria}}
\maketitle \baselineskip=11.6pt
\begin{abstract}For two bound-state equations derived as simplified
forms of the Bethe--Salpeter equation with confining interaction,
stability of all solutions is rigorously shown.\end{abstract}
\baselineskip=14pt

\section{Motivation: Instabilities of Klein-Paradox Type}The
Salpeter equation is a frequently applied three-dimensional
reduction of the Bethe--Salpeter formalism describing bound states
within quantum field theory, derived by assuming all interactions
to be instantaneous. For given interactions, encoded in its {\em
integral kernel\/}
$K(\mbox{\boldmath{$p$}},\mbox{\boldmath{$q$}})$ depending on
relative three-momenta
$\mbox{\boldmath{$p$}},\mbox{\boldmath{$q$}}$ of the bound-state
constituents, it can be regarded as an eigenvalue equation for the
Salpeter amplitude $\Phi(\mbox{\boldmath{$p$}}),$ with the mass
$M$ of the bound state as eigenvalue. For confining interactions,
however, its solutions exhibit in {\em numerical\/}
studies\cite{StabSal} certain instabilities, possibly related to
Klein's paradox, causing states to decay.

In view of this highly unsatisfactory state of the art, we began a
systematic rigorous analysis of the conditions for stability of
the energy levels predicted,~for confining interactions, within
this framework. We regard a bound state as {\em stable\/} if its
mass eigenvalue $M$ belongs to a {\em real discrete\/} spectrum
{\em bounded from below}.

On energetic grounds, any instabilities should manifest themselves
first for pseudoscalar bound states; accordingly, we focus to
fermion--antifermion bound states characterized by
spin-parity-charge conjugation assignment $J^{PC}=0^{-+}.$ This
allows to take advantage of experience gained in earlier
studies\cite{Lucha00:IBSEm=0,Lucha00:IBSE-C4,Lucha00:IBSEnzm,Lucha01:IBSEIAS}.

We analyze three-dimensional reductions of the Bethe--Salpeter
formalism for increasing complexity of the problem: the {\em
reduced\/} Salpeter
equation\cite{Lucha07:HORSE,Lucha07:StabOSS-QCD@Work07}, a
generalization thereof, proposed in Ref.~[8] and applied in
Ref.~[9], towards~exact propagators of the bound-state
constituents\cite{Lucha07:REPEHI}, and the {\em full\/} Salpeter
equation.

With Dirac couplings $\Gamma,$ assumed to be identical for both
constituents,~and related potential functions
$V_\Gamma(\mbox{\boldmath{$p$}},\mbox{\boldmath{$q$}}),$ the
action of some kernel
$K(\mbox{\boldmath{$p$}},\mbox{\boldmath{$q$}})$ on
$\Phi(\mbox{\boldmath{$p$}})$~is
$$[K(\mbox{\boldmath{$p$}},\mbox{\boldmath{$q$}})\,\Phi(\mbox{\boldmath{$q$}})]=\sum_\Gamma
V_\Gamma(\mbox{\boldmath{$p$}},\mbox{\boldmath{$q$}})\,\Gamma\,\Phi(\mbox{\boldmath{$q$}})\,\Gamma\
.$$For all interactions of harmonic-oscillator type in
configuration space, the above bound-state equations simplify to
ordinary differential equations, which may be cast into the form
of ``tractable'' eigenvalue equations for Schr\"odinger operators,
at least in the case of the {\em reduced\/} bound-state equations
studied in Secs.~\ref{Sec:RSE} and~\ref{Sec:IBSEWEP}.

Our primary tool is a (well-known) {\em theorem\/} which states
that the {\em spectrum\/} of a Schr\"odinger Hamiltonian operator
with a locally bounded positive potential $V$ rising to infinity
in all directions, $V(x)\to+\infty$ for $|x|\to\infty,$ is purely
discrete.

\section{Reduced Salpeter Equation}\label{Sec:RSE}Approximating the
propagation of both bound-state constituents by that of~free
particles of {\em constant\/} effective ``constituent'' mass $m$
yields the Salpeter equation
\begin{eqnarray}\Phi(\mbox{\boldmath{$p$}})&=&\int\frac{{\rm
d}^3q}{(2\pi)^3}\left(\frac{\Lambda^+(\mbox{\boldmath{$p$}})\,\gamma_0\,
[K(\mbox{\boldmath{$p$}},\mbox{\boldmath{$q$}})\,\Phi(\mbox{\boldmath{$q$}})]\,
\Lambda^-(\mbox{\boldmath{$p$}})\,\gamma_0}
{M-2\,E(\mbox{\boldmath{$p$}})}\right.\nonumber\\[1ex]
&&\hspace{8.7ex}\left.-\frac{\Lambda^-(\mbox{\boldmath{$p$}})\,\gamma_0\,
[K(\mbox{\boldmath{$p$}},\mbox{\boldmath{$q$}})\,\Phi(\mbox{\boldmath{$q$}})]\,
\Lambda^+(\mbox{\boldmath{$p$}})\,\gamma_0}
{M+2\,E(\mbox{\boldmath{$p$}})}\right),\label{Eq:SE-CMS}\end{eqnarray}
with {\em one-particle energy\/} $E(\mbox{\boldmath{$p$}})$ and
{\em energy projection operators\/}
$\Lambda^\pm(\mbox{\boldmath{$p$}})$
defined~by$$E(\mbox{\boldmath{$p$}})\equiv
\sqrt{\mbox{\boldmath{$p$}}^2+m^2}\ ,\qquad
\Lambda^\pm(\mbox{\boldmath{$p$}})\equiv\frac{E(\mbox{\boldmath{$p$}})
\pm\gamma_0\,(\mbox{\boldmath{$\gamma$}}\cdot\mbox{\boldmath{$p$}}+m)}
{2\,E(\mbox{\boldmath{$p$}})}\ .$$Dropping of the second term on
its RHS yields the {\em reduced Salpeter
equation\/}\cite{Henriques76}
\begin{equation}[M-2\,E(\mbox{\boldmath{$p$}})]\,\Phi(\mbox{\boldmath{$p$}})=
\int\frac{{\rm
d}^3q}{(2\pi)^3}\,\Lambda^+(\mbox{\boldmath{$p$}})\, \gamma_0\,
[K(\mbox{\boldmath{$p$}},\mbox{\boldmath{$q$}})\,\Phi(\mbox{\boldmath{$q$}})]\,
\Lambda^-(\mbox{\boldmath{$p$}})\,\gamma_0\
.\label{Eq:RSE-CMS}\end{equation}Because of its projector
structure, a pseudoscalar $\Phi(\mbox{\boldmath{$p$}})$ has just
one independent component. For a large class of kernels, all its
solutions prove to be
stable\cite{Lucha07:HORSE,Lucha07:StabOSS-QCD@Work07}.

\section{Instantaneous Bethe--Salpeter Equation with Exact
Propagators}\label{Sec:IBSEWEP}By Lorentz covariance, the exact
fermion propagator $S(p)$ is fully determined by two real
$p$-dependent (Lorentz-scalar) functions, which may be
interpreted, e.g., as this fermion's mass function $m(p^2)$ and
wave-function renormalization~$Z(p^2)$:$$S(p)=\frac{{\rm
i}\,Z(p^2)}{\not\!p-m(p^2)+{\rm i}\,\varepsilon}\
,\qquad\not\!p\equiv p^\mu\,\gamma_\mu\
,\qquad\varepsilon\downarrow0\ .$$If these propagator functions
are assumed to depend {\em approximately\/} only on the
three-momentum $\mbox{\boldmath{$p$}},$ an {\em exact-propagator
bound-state equation\/} may be found \cite{Lucha05:IBSEWEP}, from
which, for free propagators, $m(p^2)\to m,$ $Z(p^2)\to 1,$
Salpeter's equation is recovered, and which may be reduced to the
exact-propagator version of~Eq.~(\ref{Eq:RSE-CMS}):
\begin{equation}[M-2\,E(\mbox{\boldmath{$p$}})]\,\Phi(\mbox{\boldmath{$p$}})=
Z^2(\mbox{\boldmath{$p$}}^2)\int\frac{{\rm
d}^3q}{(2\pi)^3}\,\Lambda^+(\mbox{\boldmath{$p$}})\,\gamma_0\,
[K(\mbox{\boldmath{$p$}},\mbox{\boldmath{$q$}})\,\Phi(\mbox{\boldmath{$q$}})]\,
\Lambda^-(\mbox{\boldmath{$p$}})\,\gamma_0\
;\label{Eq:RIBSEWEP-CMS}\end{equation}here, one-particle energy
$E(\mbox{\boldmath{$p$}})$ and energy projection operators
$\Lambda^\pm(\mbox{\boldmath{$p$}})$ now read
$$E(\mbox{\boldmath{$p$}})\equiv
\sqrt{\mbox{\boldmath{$p$}}^2+m^2(\mbox{\boldmath{$p$}}^2)}\
,\qquad
\Lambda^\pm(\mbox{\boldmath{$p$}})\equiv\frac{E(\mbox{\boldmath{$p$}})
\pm\gamma_0\left[\mbox{\boldmath{$\gamma$}}\cdot\mbox{\boldmath{$p$}}+
m(\mbox{\boldmath{$p$}}^2)\right]}{2\,E(\mbox{\boldmath{$p$}})}\
.$$For reasonably well-behaved $m(\mbox{\boldmath{$p$}}^2)>0$ and
$Z(\mbox{\boldmath{$p$}}^2)>0,$ bound states are stable.

\section{(Full) Salpeter Equation}Trivially, similar considerations
may be applied to the full Salpeter equation (\ref{Eq:SE-CMS}).
There any analogous analysis is, however, considerably more
complicated for, at least, two reasons. On the one hand,
full-Salpeter amplitudes involve more than one independent
components. Eq.~(\ref{Eq:SE-CMS}) therefore reduces to a {\em
set\/} of second-order differential equations, equivalent to a
single differential equation of higher order. On the other hand,
although all mass eigenvalues squared $M^2$ are guaranteed to be
real, the spectrum of mass eigenvalues $M$ is in general not
necessarily real:~for the phenomenologically perhaps most relevant
interaction kernels this spectrum is a union of real opposite-sign
pairs $(M,-M)$ and imaginary points~$M=-M^\ast$.

\end{document}